\documentclass[prl,a4paper,12pt,onecolumn,superscriptaddress,amsmath,amsfonts,amssymb,preprintnumbers,citeautoscript]{revtex4}
\pdfoutput=1
\usepackage[T1]{fontenc}\usepackage[latin1]{inputenc}
\usepackage{graphicx,color,booktabs,microtype,afterpage,sidecap}
\usepackage[charter]{mathdesign}

\usepackage[colorlinks,plainpages=false,linkcolor=black,urlcolor=blue,citecolor=black,pdfpagemode=UseNone,pdfstartview=FitBH]{hyperref}

\makeatletter
\renewcommand{\@evenfoot}{\hfill\raisebox{-3em}{\bf\thepage}\hfill}
\renewcommand{\@oddfoot}{\hfill\raisebox{-3em}{\bf\thepage}\hfill}
\makeatother

\begin{document}

\title{Orbital reflectometry}

\author{Eva~Benckiser}
\affiliation{Max-Planck-Institut f\"{u}r Festk\"{o}rperforschung, Heisenbergstraße 1, 70569 Stuttgart, Germany}

\author{Maurits~W.~Haverkort}
\affiliation{Max-Planck-Institut f\"{u}r Festk\"{o}rperforschung, Heisenbergstraße 1, 70569 Stuttgart, Germany}

\author{Sebastian~Br\"{u}ck}
\affiliation{Max-Planck-Institut f\"{u}r Metallforschung, Heisenbergstr. 3, 70569 Stuttgart, Germany}
\affiliation{Experimentelle Physik 4, Physikalisches Institut, Am Hubland, 97074 W\"{u}rzburg, Germany}

\author{Eberhard~Goering}
\affiliation{Max-Planck-Institut f\"{u}r Metallforschung, Heisenbergstr. 3, 70569 Stuttgart, Germany}

\author{Sebastian~Macke}
\affiliation{Max-Planck-Institut f\"{u}r Metallforschung, Heisenbergstr. 3, 70569 Stuttgart, Germany}

\author{Alex~Fra\~{n}\'{o}}
\affiliation{Max-Planck-Institut f\"{u}r Festk\"{o}rperforschung, Heisenbergstraße 1, 70569 Stuttgart, Germany}

\author{Xiaoping~Yang}
\affiliation{Max-Planck-Institut f\"{u}r Festk\"{o}rperforschung, Heisenbergstraße 1, 70569 Stuttgart, Germany}
\affiliation{Division of Materials Science, Nanyang Technological University, 50 Nanyang Avenue, 639798 Singapore}

\author{Ole~K.~Andersen}
\affiliation{Max-Planck-Institut f\"{u}r Festk\"{o}rperforschung, Heisenbergstraße 1, 70569 Stuttgart, Germany}

\author{Georg~Cristiani}
\affiliation{Max-Planck-Institut f\"{u}r Festk\"{o}rperforschung, Heisenbergstraße 1, 70569 Stuttgart, Germany}

\author{Hanns-Ulrich~Habermeier}
\affiliation{Max-Planck-Institut f\"{u}r Festk\"{o}rperforschung, Heisenbergstraße 1, 70569 Stuttgart, Germany}

\author{Alexander~V.~Boris}
\affiliation{Max-Planck-Institut f\"{u}r Festk\"{o}rperforschung, Heisenbergstraße 1, 70569 Stuttgart, Germany}

\author{Ioannis~Zegkinoglou}
\affiliation{Max-Planck-Institut f\"{u}r Festk\"{o}rperforschung, Heisenbergstraße 1, 70569 Stuttgart, Germany}

\author{Peter~Wochner}
\affiliation{Max-Planck-Institut f\"{u}r Festk\"{o}rperforschung, Heisenbergstraße 1, 70569 Stuttgart, Germany}

\author{Heon-Jung~Kim}
\affiliation{Max-Planck-Institut f\"{u}r Festk\"{o}rperforschung, Heisenbergstraße 1, 70569 Stuttgart, Germany}
\affiliation{Department of Physics, Daegu University, Jillyang, Gyeongsan 712-714, South Korea}

\author{Vladimir~Hinkov}
\email[email: ]{v.hinkov@fkf.mpg.de}
\affiliation{Max-Planck-Institut f\"{u}r Festk\"{o}rperforschung, Heisenbergstraße 1, 70569 Stuttgart, Germany}

\author{Bernhard~Keimer}
\email[email: ]{b.keimer@fkf.mpg.de}
\affiliation{Max-Planck-Institut f\"{u}r Festk\"{o}rperforschung, Heisenbergstraße 1, 70569 Stuttgart, Germany}

\begin{abstract}
\center\bigskip\thispagestyle{plain}
\begin{minipage}{\textwidth}\textbf{The occupation of \emph{d}-orbitals controls the
magnitude and anisotropy of the inter-atomic electron transfer in
transition metal oxides and hence exerts a key influence on
their chemical bonding and physical properties.\cite{Tokura2000,Maekawa2004}
Atomic-scale modulations of the orbital occupation at surfaces and interfaces
are believed to be responsible for massive variations of the
magnetic and transport properties,\cite{Chakhalian2007,Rata2008,Jackeli2008,Tebano2008,Aruta2009,Salluzzo2009,Yu2010} but could thus far not
be probed in a quantitative manner.
Here we show that it is possible to derive quantitative, spatially resolved orbital polarization profiles from
soft x-ray reflectivity data, without resorting to model
calculations. We demonstrate that the method is sensitive enough to resolve differences
of $\mathbf{\sim 3\%}$ in the occupation of Ni \emph{e}$_\mathbf{g}$
orbitals in adjacent atomic layers of a
LaNiO$_\mathbf{3}$-LaAlO$_\mathbf{3}$ superlattice, in good
agreement with ab-initio electronic-structure calculations.
The possibility to quantitatively correlate theory and experiment on the atomic scale
opens up many new perspectives for orbital physics in \emph{d}-electron materials. }\end{minipage}
\end{abstract}

\maketitle\thispagestyle{empty}\clearpage

The electronic properties of transition metal oxides (TMOs) are determined by the interplay of the spin, charge, and
orbital degrees of freedom of the valence electrons. Progress in
understanding and predicting these properties relies on quantitative experimental
information about the spatial variation of all three observables on the atomic scale.
Powerful probes of the spin and charge densities
are already available. For instance, neutron diffractometry and reflectometry are routinely used to determine
the magnetization profiles in the bulk and near surfaces and interfaces, respectively. The valence electron charge density is more difficult to investigate, because most scattering probes couple to the total charge that is dominated by the core electrons.
Recently, however, spectroscopic methods such as electron energy-loss spectroscopy\cite{Muller2009} and
soft x-ray reflectometry\cite{Smadici2009} have yielded atomically resolved profiles of the valence electron charge.

The $d$-orbital degree of freedom is the distinguishing characteristic of TMOs compared to materials
with valence electrons in the $s$- and $p$-electron shells. In bulk TMOs, spatial alternation of the $d$-orbital occupation
(``orbital order'') is known to generate ``colossal'' variations of the
macroscopic properties,\cite{Tokura2000,Maekawa2004}
and the influence of ``orbital reconstructions'' on
the physical properties of surfaces and interfaces is currently a subject of intense investigation.\cite{Chakhalian2007,Rata2008,Jackeli2008,Tebano2008,Aruta2009,Salluzzo2009,Yu2010}
At present, however, only the spatial average of the orbital occupation can be determined
in a facile and quantitative manner, by means of x-ray linear dichroism (XLD), a method that relies
on the excitation of core electrons into the valence $d$-orbitals by
linearly polarized photons.\cite{Stoehr2006} Despite its obvious scientific interest,
quantitative experimental information about spatial variations of the $d$-orbital polarization is very limited.
Methods used to determine the amplitude of staggered orbital order
in the bulk are mostly qualitative and/or require extensive model calculations that add substantial uncertainties.\cite{Maekawa2004}
Even less information is available on orbital polarization profiles near surfaces and interfaces.
Some evidence of orbital reconstructions near single TMO interfaces has been derived
from comparison of XLD measurements in different detection modes\cite{Chakhalian2007} and from comparison to
corresponding bulk data,\cite{Tebano2008,Aruta2009,Salluzzo2009} but
these methods do not provide depth resolution.

Here we focus on TMO interfaces, which are currently in the center of a large-scale research effort
driven by prospects to control and ultimately design their electronic properties.\cite{Mannhart2010}
Since the orbital occupation determines the electronic
bandwidth and the magnetic exchange interactions at and near the interfaces, detailed experimental information on the
orbital polarization is essential for the design of suitable heterostructures.
We show that quantitative, depth-resolved profiles of the
orbital occupation in TMO heterostructures and superlattices can be
derived from linearly polarized x-ray reflectometry with photon energies near
transition-metal \textit{L}-absorption edges. The orbital occupation is extracted by the application of sum rules,
analogous to XLD, so that model calculations are not required. The method is broadly applicable to
TMO surfaces, interfaces, and multilayers and can be readily generalized to bulk diffractometry, where it will allow
quantitative measurements of staggered orbital order.

\begin{figure}[b]
\center\includegraphics[clip, width=0.6\columnwidth]{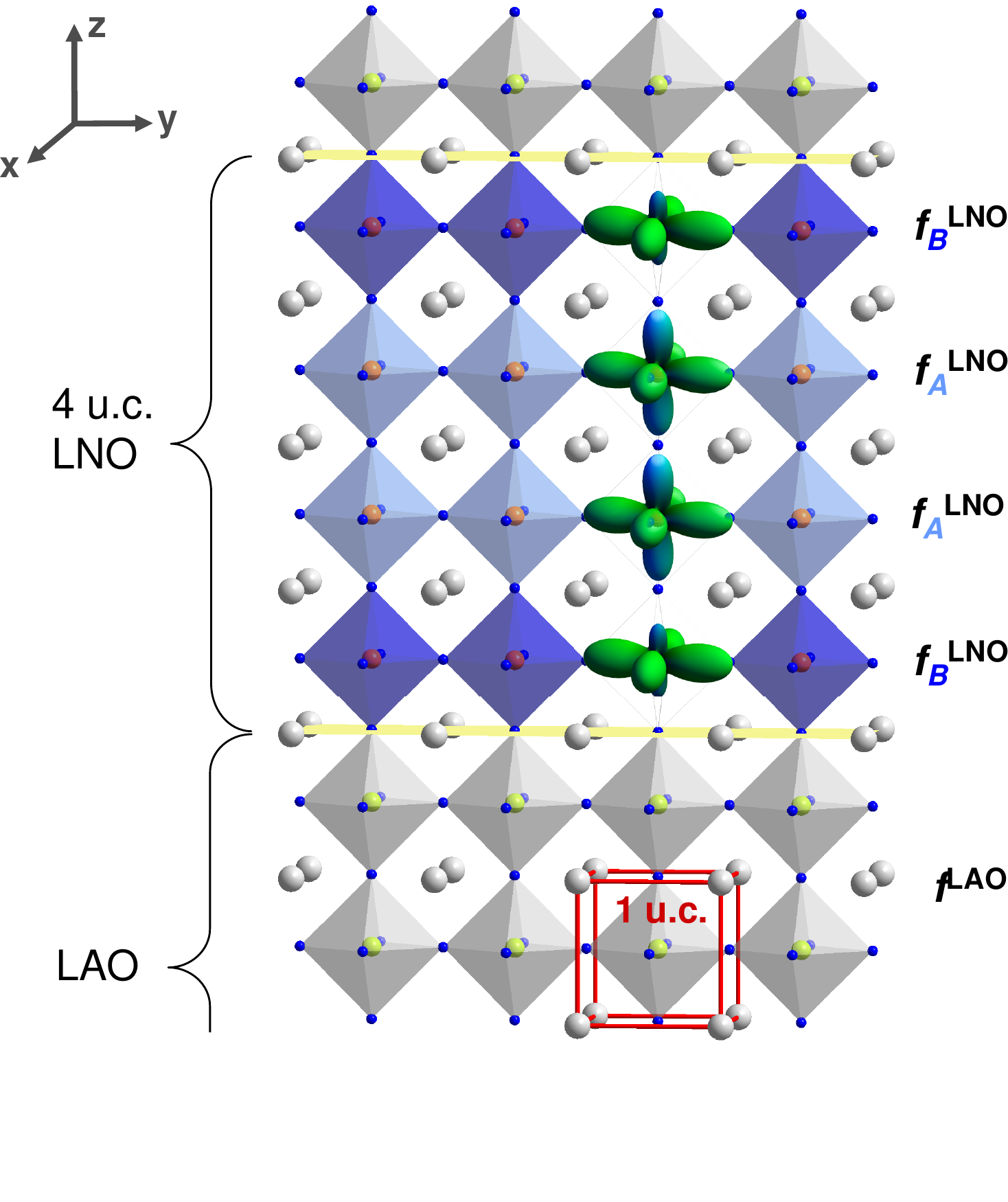}
\caption{Sketch of the LNO-LAO superlattice with layer stacks of four pseudo-cubic unit cells (u.c., see the red box), investigated in this work.
The modulation of the Ni $3d$ $e_g$ orbital occupation along the superlattice normal $z$ is depicted by a different mixture of $x^2$-$y^2$ and $3z^2$-$r^2$
orbitals and modeled with different scattering tensor $f^{\text{LNO}}_{A/B}$ (see text). The orbital occupation imbalance has been overstated for clarity.} \label{Fig1}
\end{figure}

We demonstrate the
method by investigating a superlattice comprised of the
paramagnetic metal LaNiO$_3$ (LNO) and the band insulator LaAlO$_3$
(LAO). The Ni$^{3+}$ ion in LNO has a $3d^7$ electron configuration,
and the nearly cubic crystal field of the perovskite structure
splits the atomic $3d$ orbital manifold into a lower-lying
triply-degenerate $t_{2g}$ level occupied by six electrons and a
higher-lying doubly-degenerate $e_g$ level with a single electron.
While in bulk LNO the two Ni $e_g$ orbitals (with $x^2$-$y^2$ and
$3z^2$-$r^2$ symmetry) are equally occupied, model calculations have
suggested that the $x^2$-$y^2$ (in-plane) orbital can be stabilized
by epitaxial strain and confinement in a superlattice geometry
(Fig.~\ref{Fig1}), and that the electronic structure of
superlattices with fully polarized in-plane orbitals matches the one
of the copper-oxide high-temperature
superconductors.\cite{Chaloupka2008,Hansmann2009,Millis2010} This
system is thus a prime candidate for ``orbitally engineered''
superconductivity.

We have used pulsed laser deposition to grow a
(4\,u.c.//4\,u.c.)\,$\times$\,8 LNO-LAO superlattice on a SrTiO$_3$
(STO) substrate (u.c.: pseudo-cubic unit cell; see Fig.~\ref{Fig1}).
Since the lattice constant of cubic STO is somewhat larger than the
pseudo-cubic lattice constants of LNO and LAO (mismatch
$\sim$\,2\,$\%$), the superlattice is expected to be under tensile
strain. The high crystalline quality and strain-state of the sample
were verified by x-ray diffraction and reciprocal-space mapping (see
the Supplementary Information). Transport\cite{May2009} and optical
ellipsometry measurements in combination with low-energy muon spin
rotation\cite{Boris2010} show that 4\,u.c.\ thick LNO layer stacks
in TMO superlattices are metallic and paramagnetic. The total
thickness of 247\,$\pm$\,8\,\AA\ of the superlattice was chosen to
ensure that the sample is thin enough to clearly resolve
total-thickness fringes, but thick enough to minimise contributions
of Ni$^{2+}$ that might occur in the vicinity of the polar
layer-substrate interface.\cite{Liu2010}

\begin{figure}[tb]
\includegraphics[clip, width=0.6\columnwidth]{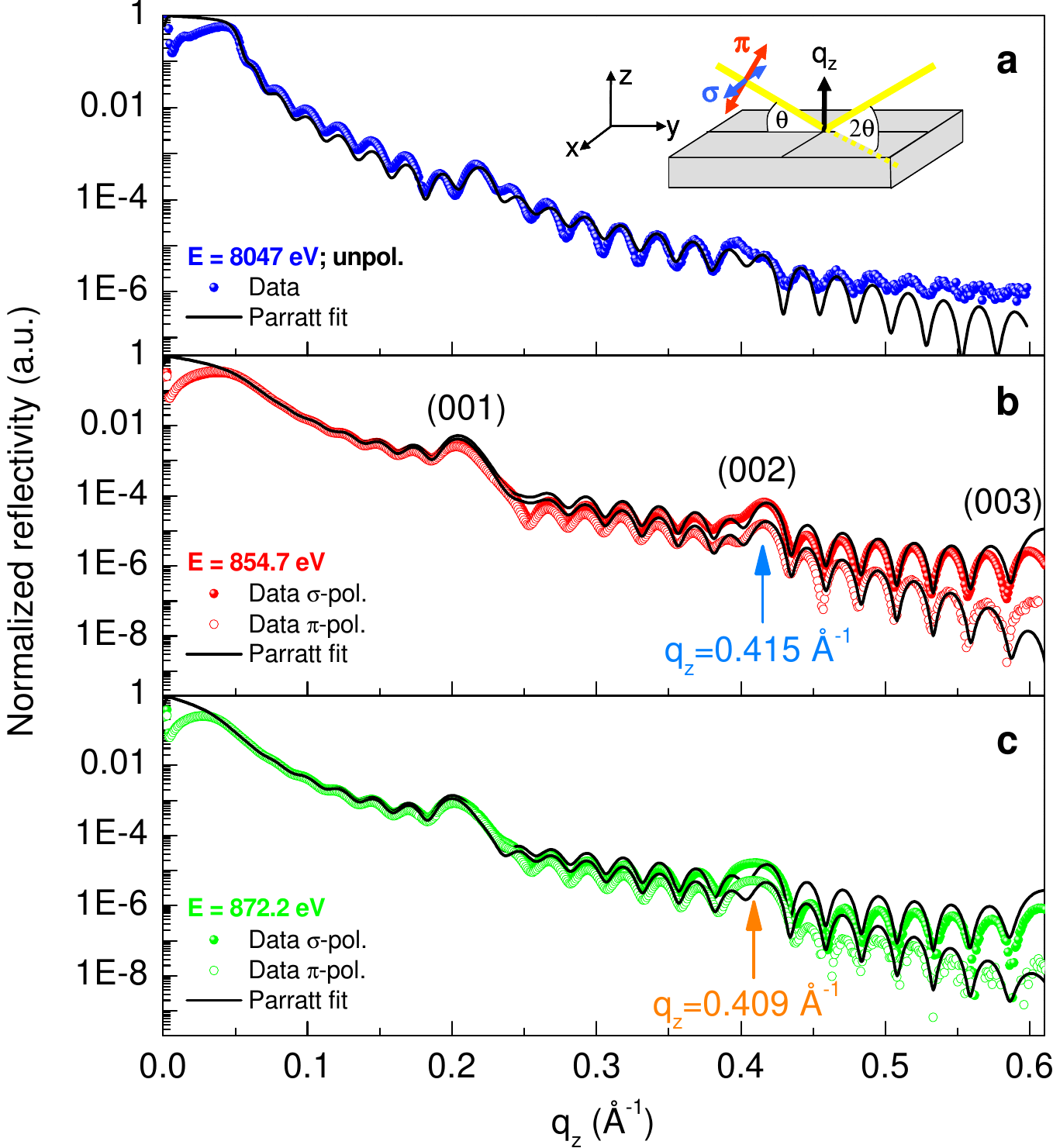}
\begin{tabular}{l c c}
\hline\hline
\footnotesize\textbf{Fit parameters}         & \footnotesize thickness (\AA)           & \footnotesize roughness (\AA)\\
\hline
\footnotesize 7 bilayer LAO/LNO (coupled) & \footnotesize 15.2/15.4   & \footnotesize 3.5/2.0   \\
\footnotesize STO substrate       &  \footnotesize $\infty$      & \footnotesize 1.7           \\
\footnotesize LNO buffer layer    &  \footnotesize 18.5          & \footnotesize 3.3          \\
\footnotesize LAO cap layer       &  \footnotesize 18.3          & \footnotesize 3.0           \\
\hline\hline
\end{tabular}\label{Tab:FitPara}
\caption{Momentum-dependent x-ray re\-flec\-tiv\-ity of the (4\,u.c.//4\,u.c.)\,$\times$\,8 LNO-LAO superlattice for
photon energies of \textbf{a}, $E$\,=\,8047\,eV (Cu $K_{\alpha}$; hard x-rays), \textbf{b}, $E$\,=\,854.7\,eV (Ni \textit{L}$_3$), and \textbf{c}, $E$\,=\,872.2\,eV (Ni \textit{L}$_2$). All data have been normalized to 1 at
$q_z$\,=\,0. The measurements in the soft x-ray range have been performed with $\sigma$ and $\pi$ polarization of the incident x-rays (see sketch in \textbf{a}).
The solid black line is the best fit to the data when using the Parratt algorithm. The fitted parameters, roughness and thickness, are summarized in the table below.} \label{Fig2}
\end{figure}

The reflectivity measurements were performed at room temperature in
specular geometry, i.e.\ the momentum transfer $q$ was parallel to
the surface normal $z$ (see the Methods section). The resulting data
for photon energy 8047\,eV, far from resonance in the hard x-ray
range (Fig.~\ref{Fig2}a) reflect the high quality of the
investigated superstructure. At energies corresponding to the Ni
\textit{L}$_{2,3}$ resonances at 854.7\,eV and 872.2\,eV, we observe
superlattice peaks up to the third order, denoted by SL (00$l$),
$l$\,=\,1, 2, 3 in Fig.~\ref{Fig2}. To obtain the layer thickness
and interface roughness of our superlattice, we fitted the
$q$-dependent data using our newly developed reflectivity analysis
program ReMagX\cite{ReMagX} in a mode based on the Parratt
formalism\cite{Parratt1954} (for details see Supplementary
Information). The resulting profiles are shown in Fig.~\ref{Fig2},
and the structural parameters are given in the table below. The best
fit is obtained when allowing the first LNO layer stack (buffer
layer) grown on STO and the last LAO layer stack (cap layer) to be
slightly thicker and different in roughness. We stress that our
structural model is simple and gives an excellent description for
the full set of data, measured for different energies and
polarization geometries.

\begin{figure}[tb]
\center\includegraphics[clip, width=0.6\columnwidth]{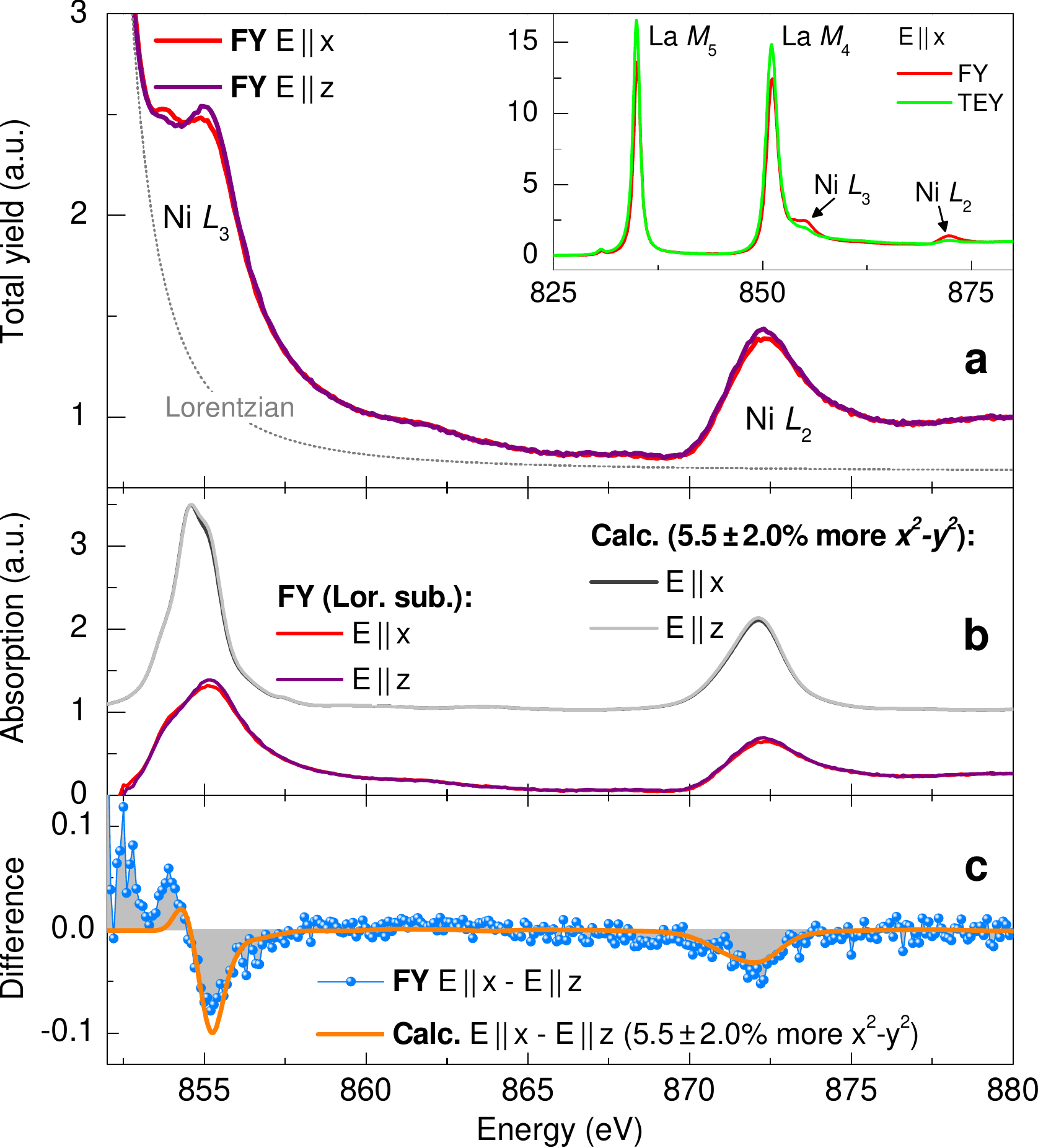}
\caption{\textbf{a}, Polarization-dependent XAS spectrum (FY) across the Ni \textit{L}$_{2,3}$ edges for $E$\,$\parallel$\,$x$ (in-plane) and $E$\,$\parallel$\,$z$
(out-of-plane) polarization. In the inset we compare TEY and FY spectra for $E$\,$\parallel$\,$x$ in the full energy range including the La \textit{M}$_{4,5}$ white lines.
\textbf{b}, Polarized FY spectra after substraction of a Lorentzian profile fitted to the La \textit{M}$_4$ line shown together with results for Ni$^{3+}$ XAS spectra with 5.5$\%$ higher $x^2$-$y^2$
occupation, obtained from the cluster calculation. \textbf{c}, difference spectra
($E$\,$\parallel$\,$x$-$E$\,$\parallel$\,$z$) calculated from the measured (blue points) and calculated data (orange line) shown in the middle panel.} \label{Fig3}
\end{figure}

The x-ray absorption spectra (XAS) measured in total-electron yield
(TEY) and fluorescence-yield (FY) modes are shown in
Fig.~\ref{Fig3}a. Due to the vicinity of the strong La
\textit{M}$_{4}$ white lines, the Ni \textit{L}$_{3}$ line is only
seen as a shoulder around 855\,eV (see inset), but the Ni
\textit{L}$_{2}$ white line is clearly observed at 872\,eV. The FY
data show a smaller intensity at the La white lines and an enhanced
contribution at the Ni lines compared to the TEY. This is explained
by self-absorption effects, limited energy resolution, different
probing depths of both methods, and the fact that the topmost layer
of the superlattice is a LAO layer. However, the dichroic difference
spectrum clearly shows dips at the Ni \textit{L}$_3$ and
\textit{L}$_2$ white line energies, which we attribute to natural
linear dichroism (Fig.~\ref{Fig3}c). We obtained nearly identical
results from FY and TEY data (not shown for clarity).

In general, natural linear dichroism reflects an anisotropy of the
charge distribution around a particular ion.\cite{Stoehr2006} To
obtain a quantitative estimate of the imbalance in $e_g$ band
occupation in our superlattice, we applied the sum rule for linear
dichroism,\cite{Thole1993,Laan1994} which relates the total
integrated intensity of the polarized spectra ($I_{E\parallel x,z}$)
to the hole occupation $\underline{n}_{3z^2-r^2}$ and
$\underline{n}_{x^2-y^2}$ in the $e_g$ orbitals:
\begin{equation}
\frac{\underline{n}_{3z^2-r^2}}{\underline{n}_{x^2-y^2}}=\frac{3I_{E\parallel z}}{4I_{E\parallel x}-I_{E\parallel z}}.
\end{equation}
By integrating the spectra in the range 853\,-877\,eV, after
subtracting the La \textit{M}$_4$ contribution estimated by a
Lorentzian profile (Fig.~\ref{Fig3}b), we obtain
$\underline{n}_{3z^2-r^2}$/$\underline{n}_{x^2-y^2}$\,=\,1.030(5),
independent of covalency. We define the orbital polarization
\begin{equation}\label{eq:OrbPolarization}
P = \frac{(n_{x^2-y^2} - n_{3z^2-r^2})}{(n_{x^2-y^2} + n_{3z^2-r^2})},
\end{equation}
with $n_{x^2-y^2}$ and $n_{3z^2-r^2}$ being the numbers of electrons
and finally obtain $P$\,$\approx$\,5\,$\%$, neglecting covalency. In
order to compare the lineshape of the polarized XAS spectra
directly, we performed a cluster calculation for a Ni$^{3+}$ ion in
an almost-cubic crystal field of six oxygen ions ($\Delta
e_g$=10\,meV) (see Fig.~\ref{Fig3}b). While the cluster calculation
is expected to give good results for localized electrons in
insulators, we expect some deviations in the calculated lineshape of
the XAS spectra in the case of our metallic nickelate layers.
Indeed, although we included reasonable line broadening (due to the
finite temperature and experimental resolution) the measured
lineshape is not fully reproduced, and we show the best result
scaled to the experimental data in such a way that the integrated
spectral weight is conserved. However, as pointed out above, the
polarization-dependence of the spectra is independent of details in
lineshape. To reproduce the measured dichroic difference,
$P$\,=\,5.5$\pm$2$\%$ higher $x^2$-$y^2$ band occupation is needed,
in good agreement with the result obtained from the sum rule. In
particular, we stress that a higher occupation of the $3z^2$-$r^2$
band would result in a sign change of the XLD signal and can be
excluded.

\begin{figure}[tb]
\center\includegraphics[clip, width=0.99\textwidth]{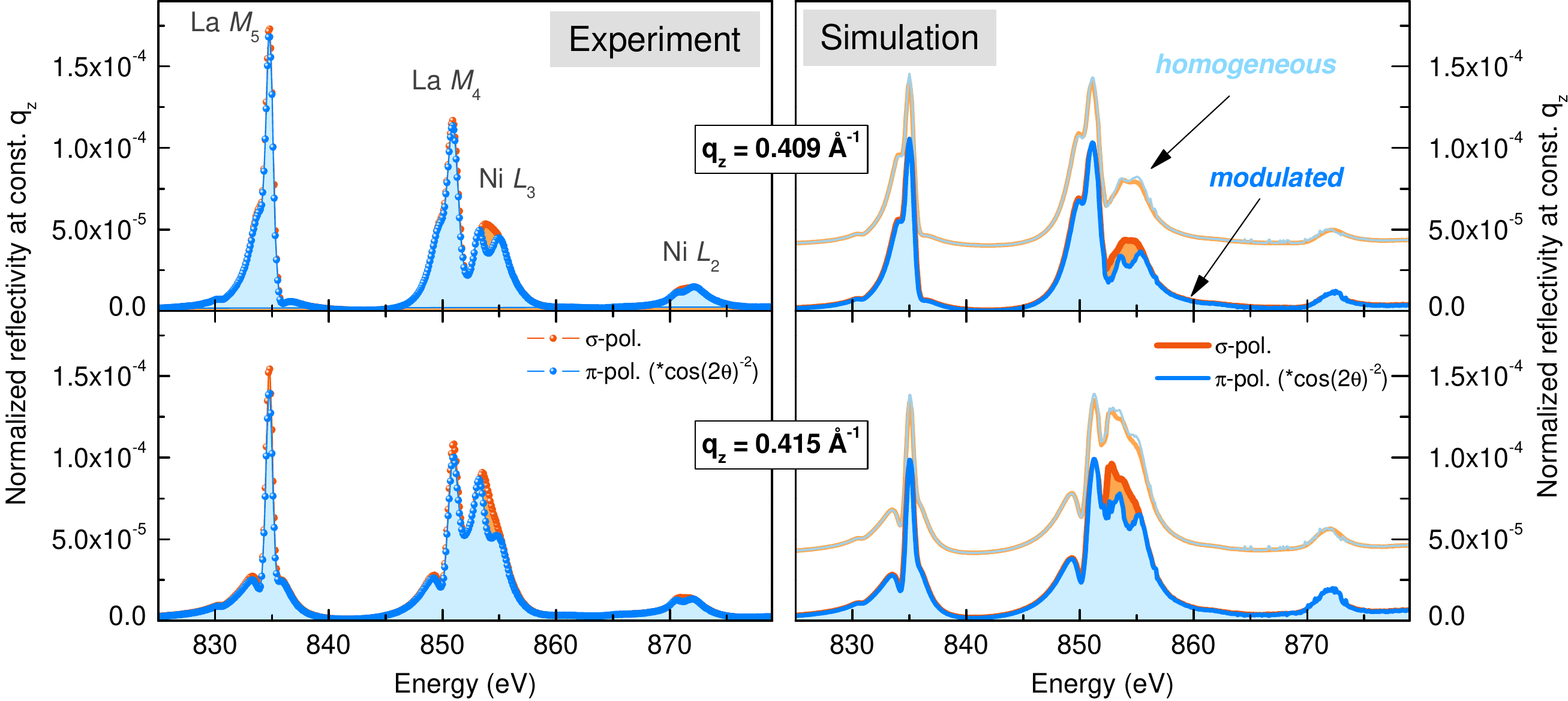}
\caption{Energy scans of the reflectivity data with constant momentum transfer $q_z$. Two $q_z$ values close to the (002) superlattice peak have been chosen from the
$q$-dependent profiles (Fig.~\ref{Fig2}): $q_z$\,=\,0.409\,\AA$^{-1}$ (top panels) and $q_z$\,=\,0.415\,\AA$^{-1}$ (bottom panels). The left panels show the polarization-dependent
experimental data. The right panels show the corresponding simulated curves for LNO layers with (i) \textit{homogeneous} orbital occupation within the LNO layer stack
(shifted by 4$\times$10$^{-5}$ for clarity) and (ii) \textit{modulated} orbital occupation of $P_B$\,=\,7$\pm$3\,$\%$ higher $x^2$-$y^2$ band occupation in the interface
layers and $P_A$\,=\,4$\pm$1\,$\%$ higher $x^2$-$y^2$ band occupation in the inner layers. Experimental data and simulations are shown for $\sigma$ and $\pi$ polarization of the
incoming photons. To correct for the difference in absolute reflected intensity for $\sigma$ and $\pi$ polarization due to the vicinity of the Brewster angle,
we multiplied the data for $\pi$ polarization with the factor $\cos{(2\theta)}^{-2}$ obtained by approximation from the Fresnel formulae.} \label{Fig4}
\end{figure}

While in XAS the averaged absorption of the LNO layers is measured,
i.e. $\text{XAS} \propto Im\left( f_A^{LNO} + f_B^{LNO}\right)$ for
LNO layer stacks composed of two inner layers with scattering factor
$f_A^{LNO}$ and two interface layers with $f_B^{LNO}$ (see
Fig.~\ref{Fig1}), the intensity of the (002) superlattice reflection
of a symmetric superlattice is mainly determined by the difference
$F^{(002)} \propto (1-i)\left(f_B^{LNO} - f_A^{LNO} \right)$. Taking
advantage of this relation, we studied the polarization dependence
of the reflected intensity across the Ni \textit{L} edge for
momentum transfers $q_z$ in the vicinity of the (002) superlattice
peak (Fig.~\ref{Fig4}). First, one notices that the lineshape of the
scattering profiles is rather different compared to the XAS profiles
and strongly depends on $q_z$. Second, there is a pronounced
polarization dependence, which only occurs at the Ni (\textit{L}$_3$
and \textit{L}$_2$) edge energies. In particular, the single-peak
profile observed for $\sigma$-polarization is split into a
double-peak profile for $\pi$-polarized photons. Although the
relative intensities change somewhat with $q_z$, nearly identical
difference spectra are observed for both $q_z$ values. A qualitative
comparison with the dichroic signal observed in the absorption
spectra already shows a clear enhancement, which indicates a
modulation of orbital occupancy \textit{within} the LNO layer stack.

In order to confirm this conclusion and extract quantitative
information about the orbital occupation, we used the newly
developed analysis tool\cite{ReMagX} to simulate the constant-$q_z$
scans (right panels of Fig.~\ref{Fig4}), based on the structural
parameters derived from $q_z$-dependent reflectometry and the
optical constants obtained from FY-XAS (see Supplementary
Information for details). To model the observed polarization
dependence, a dielectric tensor of at least tetragonal symmetry has
to be taken into account. Within the framework of the optical
approach for multilayers, we implemented formulae from the
magneto-optical formalism derived in Ref.
\onlinecite{Visnovsky2006}. We modeled our data with LNO layer
stacks split into four unit-cell thick layers,
labeled A and B in the following (Fig. 1). For each layer A and B we
assumed a dielectric tensor with tetragonal symmetry of the form
\begin{eqnarray}
\scriptsize
\hat{\varepsilon}^{\text{LNO}}_{\text{tetra A(B)}}=\left( \begin{array}{ccc}
\varepsilon^{xx}_{\text{A(B)}} & 0 & 0\\
0 & \varepsilon^{xx}_{\text{A(B)}} & 0\\
0 & 0 & \varepsilon^{zz}_{\text{A(B)}}\\
\end{array}\right)
\end{eqnarray}
with complex entries $\varepsilon^{jj}=\varepsilon^{jj}_1 +
i\varepsilon^{jj}_2$ ($j=x,z$) calculated from the optical constants
$\delta, \beta$ by
$\varepsilon^{jj}_1=(1-\delta_{jj})^2-\beta_{jj}^2$ and
$\varepsilon^{jj}_2=2(1-\delta_{jj})\beta_{jj}$. In the case of a
\textit{homogeneous} LNO stack with B and A layer having the same
dichroism, the measured averaged dichroism was taken as input,
i.e.\, $\varepsilon^{xx}$ and $\varepsilon^{zz}$ obtained from the
XAS for $E$\,$\parallel$\,$x$ and $E$\,$\parallel$\,$z$,
respectively. The simulated constant-$q_z$ scans for the
\textit{homogeneous} LNO stack cannot reproduce the large anisotropy
observed in the experiment (see light blue/orange lines in the right
panels of Fig.~\ref{Fig4}). We therefore considered the case of a
\textit{modulated} LNO stack with different $x^2$-$y^2$ band
occupation in layers A and B, keeping the averaged dichroism of
5.5$\pm$2$\%$ obtained from XLD fixed. To model the optical
constants, we parameterized them in the following way: for the cubic
case the tensor has diagonal entries
$\varepsilon_{\text{cubic}}=\frac{1}{3}(2\varepsilon^{xx}+\varepsilon^{zz})$.
Using this and the parameter $\alpha\in[0,1]$ we can write the
tensor elements:
\begin{equation}
\left(\varepsilon^{\text{LNO}}_{\text{tetra A(B)}}\right)_{jj}=(1\mp\alpha)\varepsilon^{jj}\pm\alpha\varepsilon_{\text{cubic}},
\end{equation}
with upper signs for layer A and lower ones for layer B. By varying
the parameter $\alpha$ and comparing the normalized difference
spectra of the constant-$q_z$ scans, we found the best agreement of
simulations and experiment for $\alpha$\,=\,0.30$\pm$0.05 and a
stacking sequence BAAB. This corresponds to a $P_B$\,=\,7$\pm$3$\%$
higher $x^2$-$y^2$ population in the interface layers B and a
$P_A$\,=\,4$\pm$1$\%$ higher $x^2$-$y^2$ population in the inner
layers A. There is a clear difference to the solution with stacking
ABBA (not shown for clarity).

In order to compare our results to the predictions of ab-initio
electronic-structure
calculations,\cite{Chaloupka2008,Hansmann2009,Millis2010} we have
performed LDA+\textit{U} calculations for the particular
superlattice geometry studied in this work with 4\,u.c.\ LNO (for
details see the Supplementary Information). Using equation
(\ref{eq:OrbPolarization}), we obtain an orbital polarization of
6$\%$ (3$\%$) for the outer (inner) NiO$_2$ layers B (A), in
excellent agreement with the experimental result. The correspondence between the theoretical and experimental
data also confirms that the orbital
population in our system is not significantly influenced by structural or chemical disorder.

We conclude that polarized resonant x-ray reflectometry in
combination with the analysis method we have presented here is a
powerful tool to accurately map out the orbital-occupation profile
in TMO multilayers. The quantitative correspondence between
theoretical predictions and experimental measurements we have
demonstrated opens up new perspectives for the synthesis of TMO
interfaces and superlattices with designed electronic properties, including but not
limited to the nickelate system we have investigated here. Potential
tuning parameters for a systematic design effort include epitaxial
strain from the substrate and the constituents of the superlattice,
the thicknesses of the individual layers, and the covalency of the
chemical bonds across the interface. Finally, we note that the methodology introduced here can be
readily generalized to other systems with nonuniform orbital occupation, such as
surfaces and bulk systems with staggered
orbital order, and thus has the potential to bring orbital physics in TMOs to a new level
of quantitative accuracy.

\small\noindent\smallskip

\noindent\textbf{Methods} High-quality superlattices with
atomic-scale precision have been grown by pulsed-laser deposition
from LaNiO$_3$ and LaAlO$_3$ stoichiometric targets using a KrF
excimer laser with 2\,Hz pulse rate and 1.6\,J/cm$^2$ energy
density. Both compounds were deposited in 0.5\,mbar oxygen
atmosphere at 730$^{\circ}$C and subsequently annealed in 1\,bar
oxygen atmosphere at 690$^{\circ}$C for 30\,min. Here we
investigated a (4\,u.c.//4\,u.c.)\,x\,8 LNO-LAO superlattice grown
on an atomically flat [001]-oriented single-crystalline STO
substrate. The crystallinity, superlattice structure and sharpness
of the interfaces (roughness $\leq$ 1 u.c.) were verified by
high-resolution hard x-ray  diffraction along the specular rod.
Reciprocal space mapping around the (103) STO peak shows that the
strain induced by the substrate is partially relaxed within a total
thickness of 247\AA\ (see Supplementary Information for details).

The resonant x-ray reflectivity and XAS measurements were carried
out at UE56/2-PGM1 soft x-ray beam line at BESSY II in Berlin,
Germany, using the advanced 3-axis UHV reflectometer described in
Ref.~\onlinecite{Brueck2008}. The undulator beamline supplies 90\,\%
linearly $\sigma$ and $\pi$ polarized light (for the notation see
sketch in Fig.~\ref{Fig2}a). Beamline settings have been chosen to
result in a bandwidth of 0.57\,eV at 840\,eV (optimized intensity).
The x-ray absorption have been measured with $\sigma$ and $\pi$
polarization of the incident light at an angle of incidence of
$\theta$\,=\,30$^{\circ}$. While for $\sigma$ polarization the
measured intensity corresponds directly to $E$\,$\parallel$\,$x$,
the intensity for $E$\,$\parallel$\,$z$ is deduced from
$I_z$\,=\,4/3$I_{\pi}$-1/3$I_x$, with $I_{\pi}$ being the measured
intensity with $\pi$ polarization.\cite{Wu2004} Reflected
intensities have been measured with a diode and were corrected by
the fluorescence background, measured independently with a second
diode. All intensities have been normalized by the incoming
intensity measured with a gold mesh.

Density functional calculations were performed using the Vienna
ab-initio simulation package (VASP) code. The projector
augmented-wave\cite{Bloechl1994,Kresse1999} method was used in the
framework of the generalized gradient approximation
(GGA).\cite{Perdew1996,Perdew1997,Kresse1994,Kresse1996} The orbital
occupation numbers were obtained by integrating the projected
density of state within a fixed sphere on Ni $x^2$-$y^2$ or
$3z^2$-$r^2$ orbitals (see Supplementary Information for details).

\smallskip

\bibliographystyle{naturemag}

\noindent\textbf{Supplementary Information} accompanies this paper on www.nature.com/nmat
\smallskip

\noindent\textbf{Acknowledgements} We acknowledge financial support from the DFG within the framework of the TRR80, Proj.\ C1. The authors thank G.~Khaliullin and G.~A.~Sawatzky for very fruitful discussions. We gratefully acknowledge the provision of synchrotron radiation and the assistance from W.~Mahler and B.~Zada at the UE562-PGM1 beamline at Helmholtz-Zentrum Berlin - BESSY II. We thank M.~Dudek for performing the hard x-ray reflectivity measurements, and S.~Heinze for taking the AFM image.

\smallskip

\noindent\textbf{Author contributions} E.B.\ carried out the
experiments and analyzed the data. M.W.H.\ made substantial
contributions to the data analysis and performed the cluster
calculations. S.B.\ and E.G.\ designed and set the experiment up.
S.B., E.G., and S.M.\ developed the analysis tool ReMagX. G.C.\ and
U.H.\ grew the superlattices by PLD. A.F., E.B., A.B., and P.W.\
characterized the samples by high-resolution x-ray diffraction.
X.Y.\ and O.K.A.\ performed the LDA+U calculations. I.Z.\ and
H.J.K.\ assisted in the experiments. V.H.\ worked on the data
collection and analysis. E.B., M.H., V.H., and B.K.\ wrote the
paper. V.H. and B.K. coordinated the project.

\smallskip

\noindent\textbf{Author Information} Correspondence and requests for materials should be addressed to
V.~H. (\href{mailto:v.hinkov@fkf.mpg.de}{v.hinkov@fkf.mpg.de}) or B.~K.\ (\href{mailto:v.b.keimer@fkf.mpg.de}{b.keimer@fkf.mpg.de}).

\clearpage

\bigskip

\bigskip

\bigskip

\end{document}